\documentclass[conference]{IEEEtran}
\usepackage{tikz}
\usetikzlibrary{arrows}
\usepackage[strings]{underscore}
\usepackage{subfigure}
\usepackage{graphicx}
\usepackage{epsfig}
\usepackage{amsmath}
\usepackage{amsfonts}
\usepackage{verbatim}
\usepackage[hmargin=0.585in,vmargin=0.95in]{geometry}
\usepackage{comment,array}
\usepackage{algorithm}
\usepackage{algpseudocode}
\algnewcommand{\LineComment}[1]{\State \(\triangleright\) #1}
\usepackage{balance}
\begin{document}

\renewcommand{\algorithmicrequire}{\textbf{Input:}}
\renewcommand{\algorithmicensure}{\textbf{Output:}}
\title{Cluster Aware Mobility Encounter Dataset Enlargement}
 
\author{Rajarshi Haldar$^\dag$, Salih Safa Bacanli$^\ddag$, Moayad Aloqaily$^\star$, Adel Ben Mnaouer$^\sigma$, and Damla Turgut$^\ddag$\vspace{5px} \\

    {$^\dag$University of Illinois Urbana-Champaign, Champaign, IL, USA}\\
	{$^\ddag$Department of Computer Science, University of Central Florida, Orlando, FL, USA}\\
    {$^\sigma $Canadian University of Dubai, UAE}\\

    {$^\star$Gnowit Inc., Ottawa, ON, Canada.}\\
	Emails: {$^\ddag$\{sbacanli, turgut\}@cs.ucf.edu,
	$^\star$maloqaily@ieee.org, 
	$^\dag$Rhaldar2@illinois.edu,
	$^\sigma$adel@cud.ac.ae
	}
}
\maketitle
\begin{abstract}
The recent emerging fields in data processing and manipulation has facilitated the need for synthetic data generation. This is also valid for mobility encounter dataset generation. Synthetic data generation might be useful to run research-based simulations and also create mobility encounter models. Our approach in this paper is to generate a larger dataset by using a given dataset which includes the clusters of people. Based on the cluster information, we created a framework. Using this framework, we can generate a similar dataset that is statistically similar to the input dataset.
We have compared the statistical results of our approach with the real dataset and an encounter mobility model generation technique in the literature. The results showed that the created datasets have similar statistical structure with the given dataset.
\end{abstract}

\IEEEpeerreviewmaketitle

\section{Introduction}

Realistic network simulations require the use of an acquired dataset. An acquired dataset may contain the GPS or local positions of the mobile wireless nodes or the encounter durations of the mobile nodes. Unlike the position based datasets, the  encounter based datasets simply state which two nodes encountered each other and between which timestamps. Encounter datasets hide the real location information of the nodes. In some cases, the real location of the node might be important in order to make a decision and we would be losing this piece of information. On the other hand, encounter structure contains just the needed piece of information for the mobile wireless network. For instance, node degree, inter-contact times and inter-contact durations are examples for these types of information. Since the dataset syntax is simple, the encounter dataset structure can also be used for other datasets containing social events which has start and finish time besides the actors (i.e. phone calls or continuous messaging).

Although the encounter datasets are easy to understand and simple to work on, acquiring an encounter dataset is not easy and it is mostly a laborious work. The researcher needs to provide wireless devices that can connect to each other and record the connection data. Then, the recruit4ed volunteers would need to carry these devices. After some period of time, the researcher needs to collect all the devices from the volunteers with the hope that the devices did not break down or got lost. After collecting the devices, the connectivity data from each device should be extracted and inconsistent data should be removed from the dataset. The network's representation level increases when the number of devices in the dataset increases. In this case, the researcher should have as many devices as possible resulting in additional expensive.

We hoped to generate a synthetic encounter dataset that has the same statistical properties with the input datasets in terms of distribution of the inter-contact times and contact durations. The university campus human mobility trace datasets used in the literature contains participants from different areas.  The dataset collected at University of Milano~\cite{milanodataset} contains traces from graduate students, faculty staff, and technical personnel, totaling 44 participants. 
University of St Andrews~\cite{standrewsdataset} and University of Cambridge~\cite{cambridgedataset} datasets contain traces of graduate and undergraduate students. These datasets are also relatively small in terms of encounter records and participant size.

The mobility patterns of different types of participants may not be similar. Faculty personnel may not go to as many buildings or classrooms or as frequently as the students. Graduate students might be going to one or two research centers and may not be taking as many classes as undergraduates. Creating a mobility model generally may prevent researchers from understanding the differences between these groups of people. It is still hard to create encounter mobility models of the stated groups as most of the datasets are still small to see the statistical distributions of encounter times and durations. By dividing these datasets into various groups, it becomes even harder to get a statistical distribution as the data size decreases substantially.

One of the most challenging part of creating an encounter based mobility model is that no detailed information about participants are stated either in the description or in the papers that used a specific dataset. The number of graduate students or staff forming the dataset is stated in the dataset description but examined datasets did not provide such information. The authors of this paper have contacted the faculty members who have provided the University of Milano dataset and acquired the information about which nodes are graduate student, academical or technical staff. MIT Reality dataset contains the participant information and answers of survey data for each participant; however, the dataset format is unlike the datasets from the University of Milano~\cite{milanodataset} or University of St Andrews~\cite{standrewsdataset}.

This becomes even more relevant when considering routing algorithms in Delay-Tolerant networks. For instance, social network aware routing~\cite{Misra2011} has been suggested as a viable alternative to traditional DTN-routing protocols sun as epidemic routing when network resources are limited. In social network aware routing, the nodes are socially connected and are clustered into communities.

The larger dataset can be used in Internet of Things (IoT) related applications. The opportunistic network routing algorithm can use the internet and ad hoc communication together. 
Another application case might be that the nodes may connect to internet to download some specific information about the encountered node. Based on the downloaded information, the node may decide to send all the messages or some of them to the encountered nodes. Using larger dataset that has similar statistical properties with the current dataset will be useful for testing IoT related opportunistic network routing algorithms.

The created datasets from this encounter mobility model can be used in network simulations while testing the routing techniques. In addition, understanding the human mobility may help researchers and entrepreneurs to use this information for new applications or messaging systems. The metrics about the dataset may give information about the network, therefore the structure of specific community in that environment. If we can compare mobility models of different communities on different environments, we can learn additional interesting information about the environment (e.g., people in the university campuses stop and talk to each other more frequently than urban areas).


\section{Related Work}

There exists research on simulating encounters between mobile radios. EMO~\cite{EMO} characterizes the node encounters using parameters such as inter-contact time, encounter durations, and the number of nodes encountered by a node. Using these parameters, it simulates encounters. However, it does not employ any clustering on the nodes and considers each node individually. This can be problematic if there are not sufficient number of encounters for each node, so it would be harder to synthetically generate new encounters.

Karagiannis et al.~\cite{Karagiannis} have examined the fundamental properties that determine the basic performance metrics for opportunistic communications. They have found, as an invariant property that there is a characteristic time, order of half a day, beyond which the distribution decays exponentially. Up to this value, the distribution in many cases follows a power law, as shown in recent work. This power law finding was previously used to support the hypothesis that inter-contact time has a power law tail, and that common mobility models are not adequate. More specifically, the CCDF of inter-contact time between the two nodes closely follows a power law decay up to a characteristic time; however, the decay becomes exponential beyond this.

Conan et al.~\cite{Conan} studied the distributions of inter-contact time and contact durations and concluded that each pair of nodes in a delay tolerant network does not share the same inter-contact time distribution as every other node pair. They provided the first formal analysis of the impact of heterogeneous exponential inter-contact time distributions on simple single-copy routing schemes. We have taken into account this aspect by using community detection and clustering.

There has been previous work on analyzing the mobility models on routing performance benchmarks. Thakur et al.~\cite{Thakur} found that the performance of the mobility models is not analogous to realistic trends. There were dramatic deviations from realism indicating serious flaws in existing models and their inadequacy as testbed tools for any type of performance evaluation purposes. Riascos and Mateos~\cite{riascosmateos} have analyzed the human urban mobility in New York and Tokyo based on the data harvested from Foursquare application.

Many studies have been performed to specify the nature of the inter-contact time distribution. Jahromi et al.~\cite{Jahromi} observed that aggregate contact times are best fitted with Pareto distributions, and aggregate inter-contact times followed a power law distribution to a certain point and after that followed an exponential distribution. They concluded that contact times and inter-contact times do not strictly follow a power law distribution. 

There are several advantages of social-based routing protocols, by identifying socially similar nodes, and utilizing the additional context information such as shared interests or community affiliations. Schurgot et al.~\cite{Schurgot} presented the evolution of DTN routing protocols and expressed that solutions using the social graph form a new class of routing protocols well suited for the opportunistic networks.

Bulut et al.~\cite{Bulut} noted that only few of the proposed delay tolerant network routing algorithms take into account the effect of social structure of the network on the design of the routing algorithm. It is noted in many studies that the movement of nodes in a mobile network and the interactions between these nodes are not purely random and homogeneous but rather present a mixture of homogeneous and heterogeneous behaviors. In a real mobile network, we always see grouping of nodes into communities such that the nodes within the same community behave similarly and the nodes from different communities show different behaviors. They demonstrate that the consideration of the underlying social structure of a delay tolerant network can help in designing better routing algorithms. In fact, analyzing the community structure of these networks has improved both the message delay as well as the number of message  copies formed in multi-copy based routing algorithms. 

Through comparing three different kinds of routing strategies~\cite{Gong}, the results show that the performance of routing algorithms can be improved by leveraging information obtained from social properties of the nodes. Encounter datasets are used to propose efficient routing techniques. Bacanli et al.~\cite{Bacanli-2015-GLOBECOM} proposed State-based Campus Routing (SCR) strategy by using University of Milano encounter dataset. Xu et al.~\cite{junxu} have
used the UAV to collect the information from the clusterhead nodes on the ground for animal monitoring application. 

There have been attempts at creating routing protocols that utilize social network properties. One such example is SimBetAge~\cite{Link}, a routing algorithm that takes into account temporal changes in a social network. This demonstrates that respecting the age of the node relations leads to a better performance than a binary representation.
\section{The Proposed Approach}

In our approach, we need to note that we aim to increase the dataset size. We create a dataset that has the same statistical properties as the input datasets. In order to create a larger dataset based on a given dataset, we need two input parameters beside the input dataset.

The number of clusters, the the number of nodes in each cluster, as well as the cluster association of all the nodes must be known in a given dataset. As an example, considering that the current dataset has only 15, 10, and 10 nodes in each of the corresponding three clusters, the user may wish to create a dataset that has 100 nodes in each cluster. This means that the user can request to have any arbitrary number of nodes in the generated dataset. In addition to these input values, we also need to know when to start and stop modeling the environment.

\subsection{Dataset}
We have used the Milano dataset, obtained from the CRAWDAD~\cite{milanodataset}. The encounter mobility trace of 44 participants are collected at University of Milano campus. The dataset contains encounter traces of graduate students, faculty, and technical staff. It contains a large dataset of contacts within a university campus, which has been chosen as a working context representing many other places where people usually move and meet in their professional/social life. Each node in the network is represented as a graduate student, faculty, or technical staff.

\subsection{Parameters}
For each encounter, we store the participants, duration and the inter-contact time. The inter-contact time is the duration between the end of this encounter and the beginning of the next one involving the same participants. There is no inter-contact time with respect to the last encounter.

\subsection{Procedure}
\subsubsection{Preprocessing}
To analyze the dataset, we first converted the data into a social network, where each node is a participant, and the weighted edges contain the number of contacts between the two nodes. The social network is then clustered into three categories: graduate student, faculty, or technical staff. For each pair of clusters, we store a list of the duration and inter-contact time of all encounters taking place between the nodes belonging to those two clusters. 

\if{used}
\begin{figure}
\centering
\begin{tikzpicture}[->,>=stealth',node distance=1cm,minimum size=1.5cm,
  thick,main node/.style={circle,draw}]
  \node [main node,align=center,font=\small\itshape] at ( 0,3) (1) {Graduate\\ Student};
  \node [main node,align=center,font=\small\itshape]  at ( -3,0)(2) {Faculty};
  \node [main node,align=center,font=\small\itshape] at ( 3,0)(3) {Staff};
  \path[every node/.style={font=\sffamily\small}]
    (1) edge [bend left=15] node {} (2)
    (2) edge [bend left=15] node {} (1)
    (2) edge [bend left=15] node {} (3)
    (3) edge [bend left=15] node {} (2)
    (1) edge [bend left=15] node {} (3)
    (3) edge [bend left=15] node {} (1)
    (1) edge [loop above] node {} (1)
    (2) edge [loop above] node {} (2)
    (3) edge [loop above] node {} (3);
\end{tikzpicture}
	\caption{Transition graph between groups}
    \label{fig:state}
\end{figure}
\fi

\subsubsection{Creation of synthetic network}
We generate a synthetic network that takes the total number of nodes and the total duration as the input. It randomly assigns nodes to the three clusters with the same ratio as the number of nodes in the clusters of the original dataset. 

After the creation of the nodes, we assign each of them a weight by sampling from the degree distribution of the nodes in the Milano dataset. We then randomly assign edges between each pair of clusters in the network. The number of edges between any two clusters is chosen such a way that the density of connections between the two clusters in the synthetic network follows the same trend as in the Milano dataset. Here, the density of edges is defined as the ratio of the number of edges between all the node pairs in the clusters and the total number of edges possible between the two clusters. The pair of nodes in which an edge exists is chosen randomly using the weights assigned above.

\subsubsection{Simulation of encounters}
After the nodes and the edges of the synthetic network has been created, we begin to simulate encounters. Between each node pair, we simulate the individual encounters. For each encounter, we assign a duration and an inter-contact time. These values are sampled from the distribution of duration and inter-contact time among all encounters between the two clusters the nodes belong to. We accomplish this using inverse transform sampling. Here, we are assuming that the distribution of the duration and the inter-contact time of all encounters between the two clusters are homogeneous. This process continues for each node pair until the time period exceeds the given duration input.

\subsubsection{Creation of Synthetic Dataset}
We then convert this synthetic data into an encounter dataset that has the same format as the input dataset. Each encounter is a tuple containing four values: the first node, the second node, the start time, and the end time of the encounter. The first node is chosen randomly by choosing a cluster. Afterwards, based on the density of the cluster, a random node number in that cluster is selected. Then, based on the distribution of the inter-encounter times of the two clusters of the first and second node, an inter-contact time is chosen as start of the contact time. Based on the distribution of the contact durations between the two clusters, a number is picked from the distribution as the contact duration. While selecting the next encounters, the inter-encounter time is chosen and contact start time is calculated by adding the inter-contact time with the finishing time of the last successful encounter duration. Since the encounter times and durations are created contiguously, the data will be sorted in order of start times of the contacts.

As it is shown in the algorithm~\ref{alg:enlarger}, the ICDBetween and ICTBetween functions are taking the two clusters as parameters. These functions return a random number based on the distribution of inter-contact times and inter-encounter times between the two cluster inputs. At the end, the algorithm returns a set of encounter entries. If these entries are sorted based on their contact start times, the dataset would be usable by simulators and also it would be easier to analyze and follow chronologically.

\begin{algorithm*}
\caption{Dataset Enlarger}
\label{alg:enlarger}
\begin{algorithmic}[1]
\Require{$inputdata$:Encounter dataset $ClusterNodes$:list $MaxTime$:number}
\Ensure{Generated Encounter Dataset as list}
\Function{EnlargeDataset}{ $datainput,ClusterNodes,maxtime$ }
\State{$GeneratedEncounters\gets \{\}$}
\LineComment{ClusterNodes is an array. Length of ClusterNodes is number of clusters.}
\LineComment {Each element contains the number of nodes in that cluster}
\LineComment {MaxTime is maximum number of seconds}
\For{$i \gets 1 $ to $length(ClusterNodes)$}   
\State{$numberOfNodesInCluster$ $\gets$ $ClusterNodes_{i}$}
\For{$node1 \gets 1 $ to $numberOfNodesInCluster$}   
\State $node2Cluster$ $\gets$ random number between [0, length of $ClusterNodes$)
\State $node2$ $\gets$ random number between [0, ($ClusterNodes_{node2Cluster-1}$))
\State $contactEnd \gets 0$
\While{$contactEnd < MaxTime$}
\State $interContactDuration \gets ICDBetween(node1Cluster,node2Cluster,inputdata)$
\State $interContactTime \gets ICTBetween(node1Cluster,node2Cluster,inputdata)$
\LineComment{Curve fitting is done while getting ICT and ICD between clusters}
\State $contactStart \gets contactEnd + interContactTime$
\State $contactEnd \gets interContactTime + interContactDuration$
\State $GeneratedEncounters \gets GeneratedEncounters \cup \{node1,node2,contactStart,contactEnd\}$
\EndWhile
\EndFor
\EndFor
\State \textbf{return} $GeneratedEncounters$
\Comment {The result can be sorted by contactStart times}
\EndFunction
\end{algorithmic}
\end{algorithm*}

We need to note that, while creating a synthetic dataset, we have used exactly the same parameters of the real dataset. In other words, the number of nodes in each cluster for the synthetic dataset is the same as the number of nodes in each cluster within the real dataset. The maximum data collection time also matches the duration of the real dataset. 
\section{Results}

We have compared the statistical distributions of inter-contact time and contact durations of our generated datasets. The datasets generated by EMO~\cite{EMO} using Milano dataset and the real Milano dataset. \setlength\belowcaptionskip{-5ex}

\begin{figure}[!htp]
  \includegraphics[width=\linewidth]{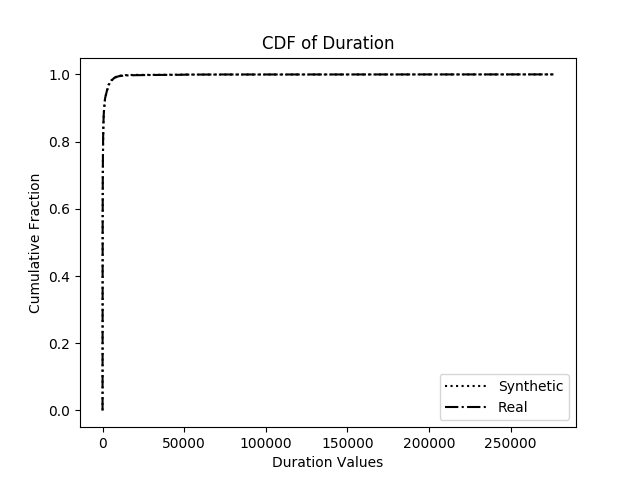}
  \caption{CDF of Inter Contact Durations of first 250000 seconds }
  \label{fig:cdfsyntheticDur}
\end{figure}
\vspace*{-5mm}
\begin{figure}[h]
  \includegraphics[width=\linewidth]{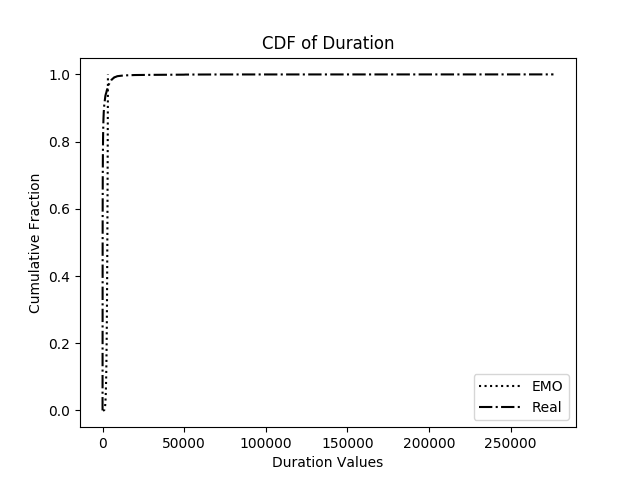}
  \caption{CDF of Inter Contact Durations of first 250000 seconds }
  \label{fig:cdfemoDur}
\end{figure}

Figure~\ref{fig:cdfsyntheticDur} and Figure~\ref{fig:cdfemoDur} show the cumulative distribution functions of inter-contact durations comparing our synthetic model and EMO. Both figures show the durations of first 250000 seconds. Our synthetic models seem to be fitting better than EMO. We only showed the results showing less than 250000 seconds as the difference is most visible when the graphs are showing results less than 250000 seconds. \setlength\belowcaptionskip{-2ex}

\begin{figure}[h]
  \includegraphics[width=\linewidth]{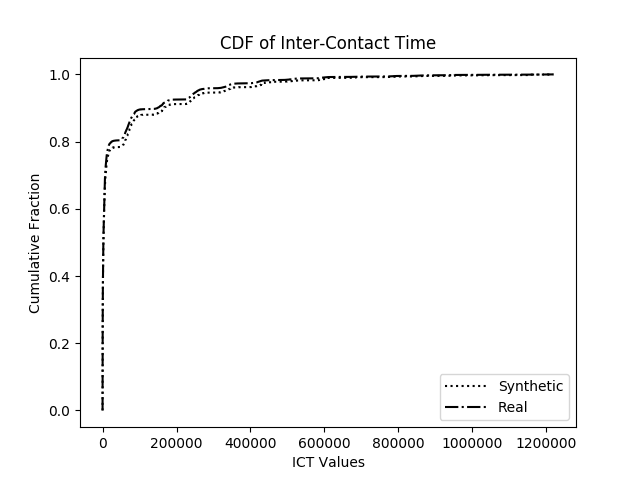}
  \caption{CDF of Inter-contact Times of first 1200000 seconds}
  \label{fig:cdfsyntheticICT}
\end{figure}

\begin{figure}[h]
  \includegraphics[width=\linewidth]{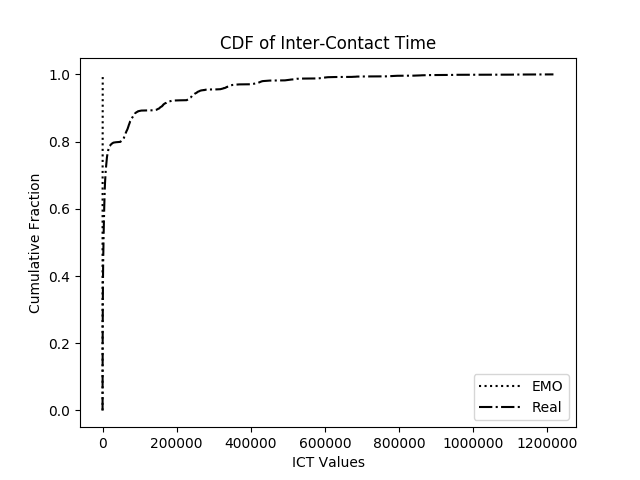}
  \caption{CDF of Inter-contact Times of first 1200000 seconds}
  \label{fig:cdfemoICT}
\end{figure}

EMO model is developed mainly to create a general encounter mobility model. EMO assumes that all encounter times and duration patterns are homogeneous and it does not take into account any cluster or category of people. Our model on the other hand is aware of the clusters of people in a dataset, achieving improved results. The statistical distribution results were as expected.\setlength\belowcaptionskip{-2ex}
\section{Conclusion}

We have modelled the mobility of graduate students, faculty, and technical staff on a university campus. By doing that we have created a framework where we can generate synthetic dataset for encounters. The created mobility model is compared with EMO~\cite{EMO} and the real dataset. The results show that the mobility model gives similar statistical distributions with real dataset and show more realistic results than EMO~\cite{EMO}.

This framework can be potentially used by the researchers to predict the next inter-contact time for a person on campus or average contact duration for that person. The datasets that contain encounter mobility in the literature are either recorded for short amount of time or the number and variety of the participants are a few. This framework also allows researchers to generate synthetic encounter mobility dataset for use in network analysis or routing protocols. For instance, the simulations on larger synthetic dataset can be leveraged to assess the efficiency of the opportunistic routing techniques.

\bibliographystyle{IEEEtran}
\bibliography{synEnc}

\end{document}